\documentclass[12pt]{iopart}
\usepackage{iopams}
\usepackage{amsbsy}

\usepackage{mathrsfs}
\usepackage{dsfont}
\usepackage{graphicx}
\usepackage{bm}
\usepackage{bbm}

\usepackage{xcolor}
\newcommand{\mvec}[1]{\bm{#1}}

\newcommand{\ket}[1]{\left | #1 \right\rangle}

\newcommand{\eqref}[1]{(\ref{#1})}
\newcommand{\betac}{\beta_\mathrm{c}}
\newcommand{\hc}{h_\mathrm{c}}

\begin{document}

\title{Finite-Temperature Fidelity-Metric Approach to the Lipkin-Meshkov-Glick Model}

\date{\today}

\author{Daniel D.\ Scherer,$^1$ Cord A.\ M\"uller$^2$ and Michael Kastner$^3$}

\address{$^1$ Theoretisch-Physikalisches Institut, Friedrich-Schiller-Universit\"at Jena, 07743 Jena, Germany}
\address{$^2$ Physikalisches Institut, Universit\"at Bayreuth, 95440 Bayreuth, Germany}
\address{$^3$ National Institute for Theoretical Physics (NITheP), Stellenbosch 7600, South Africa}

\ead{\mailto{daniel.scherer@uni-jena.de}, \mailto{cord.mueller@uni-bayreuth.de}, \mailto{kastner@sun.ac.za}}

\begin{abstract}
The fidelity metric has recently been proposed as a useful and elegant approach to identify and characterize both quantum and classical phase transitions. We study this metric on the manifold of thermal states for the Lipkin-Meshkov-Glick (LMG) model. For the isotropic LMG model, we find that the metric reduces to a Fisher-Rao metric, reflecting an underlying classical probability distribution. Furthermore, this metric can be expressed in terms of derivatives of the free energy, indicating a relation to Ruppeiner geometry. This allows us to obtain exact expressions for the (suitably rescaled) metric in the thermodynamic limit. The phase transition of the isotropic LMG model is signalled by a degeneracy of this (improper) metric in the paramagnetic phase. Due to the integrability of the isotropic LMG model, ground state level crossings occur, leading to an ill-defined fidelity metric at zero temperature. 
\end{abstract}


\section{Introduction}\label{sec:1}

Beginning with ground-state overlap studies of Zanardi and Paunkovi\'c \cite{Zanardi2006}, the fidelity of quantum states has recently been used for investigating classical as well as quantum critical behaviour in various 
systems. The motivating idea behind this approach is simple, yet extremely plausible: The properties of different macroscopic phases of matter should be encoded in the structure of rather distinct quantum states. Hence, a suitable metric that can quantify how ``different'' two given quantum states are should be able to capture some signature of a phase transition (see \cite{Gu2008} for a recent review of these and related ideas). 

The appeal of this approach lies in the fact that it is related to geometric structures inherent to the state space of the given quantum system itself. This was already pointed out in \cite{Zanardi2007b} for the case of pure quantum states, and a generalization of fidelity to finite temperatures was discussed in \cite{Zanardi2007e}. Fidelity itself and the corresponding geometric quantities might thus serve as ``universal order parameters'' that reveal signatures of criticality at zero as well as finite temperatures. A related approach, proposing the use of the so-called fidelity susceptibility in order to identify and characterize quantum phase transitions, has recently been put forward by You \textit{et al.} \cite{You2007}. 

A further interesting feature of the fidelity-metric approach lies in the fact that it also applies to non-standard (quantum) phase transitions, like topologically ordered phases \cite{XGWen}. For such transitions, no symmetry breaking principles are at work, and no local order parameter can be defined. For current results on fidelity and fidelity metric approaches to topological order, see \cite{Garnerone2009, Abasto2008}. 

In the present article, we study the phase transition of the Lipkin-Meshkov-Glick (LMG) model within the fidelity-metric approach \cite{Lipkin1965}. Originally, this model was proposed to describe excitations in simple atomic nuclei. In its spin-$1/2$ representation, it can be regarded as a quantum $XY$ model with infinite-ranged ferromagnetic exchange interactions, where every spin is subject to an external transverse magnetic field $h$. This model shows a continuous phase transition from a symmetry-breaking, ferromagnetically ordered phase to a phase that is spin-polarized for zero temperature and high fields and crosses over continuously to a paramagnet at zero field and high temperature. We mostly study the isotropic case, being rotationally symmetric in the $(x,y)$-plane. This case is somewhat special due to the fact that the Hamiltonian consists of mutually commuting terms, and no ``competition'' between noncommuting terms (regarding e.g.\ symmetry) can arise. Our aim is then to obtain the Riemannian metric tensor field related to fidelity, defined on the model's quantum state space. As expected, we find in this metric a signature of the phase transition. The peculiarities of the isotropic LMG model lead to a number of remarkable properties of the metric: First, as a consequence of exact ground-state level crossings, the metric is not well defined on the ground state manifold, i.e., at zero temperature. Second, for finite temperatures, we find a very pronounced signature at the phase boundary, with a well-defined Riemannian metric for the ferromagnetic phase, and a degenerate tensor field (not being a proper Riemannian metric) for the paramagnetic phase. Third, the metric components can be expressed entirely in terms of derivatives of the free energy, suggesting a close relation to Ruppeiner geometry \cite{Ruppeiner1995}. These features should disappear for the anisotropic case, i.e., as soon as a noncommuting term is added to the Hamiltonian. 

Studies of the phase transition of the LMG model within the framework of fidelity, fidelity susceptibility and related concepts have been reported previously. In \cite{Quan2008}, fidelity was used basically as an alternative means to obtain the phase diagram, whereas in \cite{Kwok2007} the fidelity susceptibility and its scaling behaviour were studied (see also \cite{Kwok2008} and \cite{Ma2008} for related work). Yet, to our knowledge, the explicit calculation of the associated metric tensor field at finite temperatures is novel. 

The article is structured as follows: In sections \ref{sec:2} and \ref{sec:3} we give an overview of quantum state space and its underlying geometric structures. This will lead us to the concepts of Fubini-Study geometry in the case of pure states and Bures geometry for mixed states. In section \ref{sec:4} we introduce the isotropic LMG model, its simple solution in terms of angular momentum states, and its exact thermodynamic solution. section \ref{sec:5} is devoted to the computation of the fidelity metric induced on thermal submanifolds and the Ricci scalar. The Fubini-Study limit is discussed in section \ref{sec:6}, and remarks on the anisotropic LMG model can be found in section \ref{sec:7}. A discussion of the results and an outlook on future work is given in section \ref{sec:8}.
 
\section{Fubini-Study Geometry on Quantum State Space $\mathcal{P}(\mathscr{H})$}\label{sec:2}

In this section we introduce a Riemannian metric on quantum state space which serves as a measure of distinguishability of quantum states. As a first step, following \cite{Pol}, we will introduce quantum state space as a base manifold of a certain fiber bundle. Then there exists a very natural (from a mathematical point of view) way to derive a metric on quantum state space from the scalar product on Hilbert space. Remarkably, this metric has an information-geometric interpretation, rendering it a useful measure of distinguishability of quantum states. 
 
Consider a quantum system defined on a Hilbert space $\mathscr{H}$. We denote by
\begin{equation}
S(\mathscr{H}) \equiv
\bigl\{|\psi\rangle\in\mathscr{H}\,\big|\,\langle\psi|\psi\rangle = 1\bigr\}\subset\mathscr{H}
\end{equation}
the subset of normalized Hilbert space vectors. Then it is well-known that the relevant physical information is contained in the transition probabilities $|\langle\psi|\varphi\rangle|^{2}$, where $|\psi\rangle,|\varphi\rangle\in S(\mathscr{H})$. However, $S(\mathscr{H})$ contains redundant state vectors, and therefore is not what we would like to call the quantum state space: For a phase-shifted Hilbert-space vector
\begin{equation}\label{U1}
|\psi^{\prime}\rangle\equiv\mathrm{e}^{\mathrm{i}\theta}|\psi\rangle, 
\end{equation}
it is obvious that $|\langle\psi^{\prime}|\varphi\rangle|^{2}=|\langle\psi|\varphi\rangle|^{2}$, and $|\psi\rangle$ and $|\psi^\prime\rangle$ cannot be distinguished by measuring expectation values of any observable acting on $\mathscr{H}$ alone. Putting it less mundane, the invariance of transition probabilities under these $U(1)$ transformations induces an equivalence relation $|\psi\rangle\sim|\psi^{\prime}\rangle$ on $S(\mathscr{H})$. We denote by $[\psi]\in\mathcal{P}(\mathscr{H})$ the corresponding equivalence classes, where the projective Hilbert space $\mathcal{P}(\mathscr{H})$ is the space of equivalence classes. The projective Hilbert space now defines our first version of a quantum state space. Note that, for finite-dimensional Hilbert spaces $\mathscr{H}=\mathds{C}^N$, the projective Hilbert space $\mathcal{P}(\mathscr{H})\cong{\mathds{C}P}^{N-1}$ is a complex projective space, which is well-studied in geometry. The projection mapping 
\begin{equation}
\pi:S(\mathscr{H})\rightarrow\mathcal{P}(\mathscr{H}),\qquad |\psi\rangle\mapsto[\psi],
\end{equation}
allows for a fiber bundle interpretation: $S(\mathscr{H})\stackrel{\pi}\rightarrow\mathcal{P}(\mathscr{H})$ is a principal fiber bundle with structure group $U(1)$ and quantum state space $\mathcal{P}(\mathscr{H})$ as its base space. 
The fibers $\pi^{-1}([\psi])$ are one-dimensional subspaces of $\mathscr{H}$ and are themselves isomorphic to $U(1)$. Note, that $\mathcal{P}(\mathscr{H})$ is isomorphic to the space of one-dimensional projectors of the form $|\psi\rangle\langle\psi|$.

The Hilbert space $\mathscr{H}$ possesses a geometric structure that originates from its scalar product in a straightforward way, 
\begin{equation}\label{eq:1} 
\langle\psi|\varphi\rangle \equiv G(\psi, \varphi) +\mathrm{i}\Omega(\psi,\varphi).
\end{equation}
Here, $G(\psi, \varphi)$ and $\Omega(\psi,\varphi)$ are defined as the real, respectively imaginary, part of $\langle\psi|\varphi\rangle$. $G:\mathscr{H}\times\mathscr{H}\rightarrow\mathds{R}$ is a bilinear, non-degenerate and symmetric map. Due to linearity of the Hilbert space $\mathscr{H}$, we can identify its tangent space $T_{\psi}\mathscr{H}$ at a point $|\psi\rangle$ with $\mathscr{H}$ itself, $T_{\psi}\mathscr{H}\cong\mathscr{H}$. Thus, $G$ can also be seen as a mapping from $T\mathscr{H}\times T\mathscr{H}$ to the reals, and indeed defines a Riemannian structure on the Hilbert space $\mathscr{H}$. Similarly, $\Omega:\mathscr{H}\times\mathscr{H}\rightarrow\mathds{R}$ defines a symplectic form on $\mathscr{H}$, and together with $G$ it endows $\mathscr{H}$ with a K\"{a}hlerian structure. The interested reader can find more information on these geometric structures and their implications for quantum mechanics in references \cite{Pol,AshSchil}. In the present article, we will be concerned exclusively with the properties of the Riemannian metric $G$. 

Our next aim is to carry over the Riemannian structure from $S(\mathscr{H})$ to the quantum state space $\mathcal{P}(\mathscr{H})$. Clearly, the Riemannian metric defined in \eqref{eq:1} is not invariant under the $U(1)$ phase rotation \eqref{U1}. But the metric structure in the projective space of equivalence classes cannot depend on these phases and should be defined accordingly. Here the bundle structure $S(\mathscr{H})\stackrel{\pi}\rightarrow\mathcal{P}(\mathscr{H})$ comes in handy. A connection on a fiber bundle introduces the notions of vertical ($\in V_{\psi}$) and horizontal vectors ($\in H_{\psi}$). Vertical vectors ``point along'' the fiber direction and are elements of the tangent spaces to the points in $\pi^{-1}([\psi])$. So given a curve $(-\epsilon,+\epsilon)\ni t\mapsto\mathrm{e}^{\mathrm{i}\theta(t)}\ket{\psi}$, $\theta(0)=0$ along the fiber $\pi^{-1}([\psi])$, the tangent vector 
$\frac{\mathrm{d}}{\mathrm{d}t}(\mathrm{e}^{\mathrm{i}\theta(t)}\ket{\psi})|_{t=0}=\mathrm{i}\dot\theta(0)\ket{\psi}
$
at the point $\ket{\psi}$
spans the vertical vector space 
$V_{\psi}
\cong\mathrm{i}\mathds{R}$.
To describe the connection in terms of a 1-form, we can naturally make use of the Hilbert-space scalar product. Take as this natural connection $\langle\psi|\,\cdot\,\rangle:T_{\psi}S(\mathscr{H})\rightarrow \mathds{C}$. Then the horizontal tangent space at a point $|\psi\rangle\in S(\mathscr{H})$ is given by those vectors which are mapped to zero by this connection, 
\begin{equation}\label{eq:2} 
H_{\psi}\equiv\bigl\{|\varphi\rangle\in T_{\psi}S(\mathscr{H})\big|\langle\psi|\varphi\rangle=0\bigr\}.
\end{equation} 
This is precisely the orthogonal complement to $|\psi\rangle$ in $S(\mathscr{H})$, yielding the decomposition
\begin{equation}
T_{\psi}S(\mathscr{H})=H_{\psi}\oplus V_{\psi}
\end{equation}
of tangent spaces of $S(\mathscr{H})$. An element $|\varphi^{H}\rangle\in H_{\psi}$ can now be written as
\begin{equation}
|\varphi^{H}\rangle=|\varphi\rangle-\langle\psi|\varphi\rangle|\psi\rangle.
\end{equation}
This enables us to define a bilinear mapping
\begin{equation}
\langle\cdot|\cdot\rangle_{[\psi]}:T_{[\psi]}\mathcal{P}(\mathscr{H})\times T_{[\psi]}\mathcal{P}(\mathscr{H})\rightarrow\mathds{C}
\end{equation}
on the tangent spaces $T_{[\psi]}\mathcal{P}(\mathscr{H})$ of $\mathcal{P}(\mathscr{H})$ as
\begin{equation}\label{eq:3}
\langle P_{1}|P_{2}\rangle_{[\psi]}\equiv\langle\varphi_{1}^{H}|\varphi_{2}^{H}\rangle.
\end{equation}
Here, $|\varphi_{1}^{H}\rangle$, $|\varphi_{2}^{H}\rangle$ are vectors which are pushed forward to $P_{1},P_{2}\in T_{[\psi]}\mathcal{P}(\mathscr{H})$ by the tangent projection $\pi_{*}:T_{\psi}S(\mathscr{H})\rightarrow T_{[\psi]}\mathcal{P}(\mathscr{H})$    
that gives the tangent vectors to the projected curves $\pi(\ket{\varphi(t)})$ in the base space $\mathcal{P}(\mathscr{H})$. 
Eq. \eqref{eq:3} can also be written as 
\begin{equation}\label{eq:4} 
\langle P_{1} |
P_{2}\rangle_{[\psi]}=\langle\varphi_{1}|\varphi_{2}\rangle-\langle\varphi_{1}|\psi\rangle\langle\psi|\varphi_{2}\rangle.
\end{equation}
This object is often referred to as the quantum geometric tensor \cite{Provost1980}. By construction, it is invariant under the $U(1)$ transformation introduced above.

Taking the real part on both sides of equation \eqref{eq:4}, we can now define
\begin{equation}\label{eq:5}
g(P_{1}, P_{2})\equiv \Re\bigl\{\langle P_{1}| P_{2}\rangle_{[\psi]}\bigr\}=G(\varphi_{1}^{H},\varphi_{2}^{H}).
\end{equation}
as a Riemannian metric on the projective Hilbert space. For explicit calculations, it proves useful to rewrite $g$ by employing a local section $\mathcal{P}(\mathscr{H})\rightarrow S(\mathscr{H})$, $[\psi]\mapsto |\psi\rangle$. This section induces a push forward 
\begin{equation}
T_{[\psi]}\mathcal{P}(\mathscr{H})\rightarrow T_{\psi}S(\mathscr{H}),\qquad P\mapsto|\mathrm{d}\psi(P)\rangle,
\end{equation}
which allows us to write $|\varphi\rangle=|\mathrm{d}\psi(P)\rangle$. Finally we obtain
\begin{equation}\label{eq:6} 
g(P_{1},P_{2})\!=\!\Re\biggl\{\!
\langle\mathrm{d}\psi(P_{1})|\mathrm{d}\psi(P_{2})\rangle-\langle\mathrm{d}\psi(P_{1})|\psi\rangle\langle\psi|\mathrm{d}\psi(P_{2})\rangle
\! \biggr\}. 
\end{equation}
This Riemannian metric is usually called Fubini-Study metric. Remarkably, one finds that the distance corresponding to this metric is a distance in the information-geometric sense, telling how `difficult' it is to distinguish between certain states by means of ideal measurements (see \cite{Bengtsson} for details). This metric is useful for studying quantum phase transitions at zero temperature, where only pure quantum states need to be considered. 

\section{Bures Geometry on Quantum State Space $\mathcal{M}$}\label{sec:3}

For the study of thermal phase transitions, we have to extend the fidelity metric to mixed states, i.e., to the space of density operators. This formalism is mostly due to Uhlmann \cite{Uhlmann1986}; the presentation in this paper mainly follows reference \cite{Pol}. 

Let $\mathcal{M}$ denote the set of density operators, which defines our second version of a quantum state space. First note that $\mathcal{P}(\mathscr{H})\subset\mathcal{M}$, since we have an isomorphism $[\psi]\mapsto|\psi\rangle\langle\psi|$ for pure states. For mixed states, one can identify a fiber bundle structure by observing that any density operator can be written $\rho=WW^{\dagger}$. The so-called purification $W\in S(\mathscr{H}^{\mathrm{HS}})$ is an element of 
\begin{equation}\label{eq:8} 
S(\mathscr{H}^{\mathrm{HS}}) \equiv \bigl\{W\in\mathscr{H}^{\mathrm{HS}}\,\big|\,\|W\|_{\mathrm{HS}}=1\bigr\}, 
\end{equation}
the Hilbert-Schmidt space of bounded operators $W:\mathscr{H}\to\mathscr{H}$ with unit norm $\|W\|_{\mathrm{HS}}\equiv\sqrt{\langle W,W\rangle_{\mathrm{HS}}}$ that is derived from the scalar product 
\begin{equation}
\langle W_{1}, W_{2}\rangle_{\mathrm{HS}}\equiv\mathrm{tr}\,W_{1}^{\dagger}W_{2}^{\phantom{\dagger}}.
\end{equation}
Now, what is this construction good for, and where is the bundle? 
The purification of a given density operator is not unique since, if $W:\mathscr{H}\to\mathscr{H}$ defines a purification, then $WV$ with $V\in U(\mathscr{H})$ purifies $\rho$ as well. Here, $U(\mathscr{H})$ denotes the group of unitary operators acting on the Hilbert space $\mathscr{H}$. So instead of considering equivalence classes over just the pure-state phases $U(1)$, 
we now introduce a projection mapping $\pi:S(\mathscr{H}^{\mathrm{HS}}) \rightarrow\mathcal{M}$ by $W\mapsto\rho=WW^{\dagger}$. 

There is still a slightly technical obstruction to obtaining a well defined $U(\mathscr{H})$ bundle. For general density operators, the ``fibers'' $\pi^{-1}(\rho)$ need not be isomorphic to each other and $U(\mathscr{H})$. This can be seen as follows. A general density matrix is by definition a positive operator and can accordingly have null eigenvalues. Consequently, an operator $W$ 
projected to a given $\rho$ is not necessarily of full rank. Moreover, if $W_{1}$ and $W_{2}$ are projected to $\rho_{1}$ and $\rho_{2}$, respectively, they can differ in rank. Hence, in the presence of null eigenvalues of $\rho$, we can expect a one-to-one correspondence for the elements of $\pi^{-1}(\rho)$ only to a {\em subgroup}\/ of $U(\mathscr{H})$. Therefore, in order to obtain a well defined $U(\mathscr{H})$-bundle with all fibers isomorphic to $U(\mathscr{H})$, we need to restrict the base space to strictly positive operators (only non-null eigenvalues), 
\begin{equation}\label{eq:9}
\mathcal{M}^{+}\equiv\bigl\{\rho\in\mathcal{M}\,\big|\,\rho>0\bigr\}.
\end{equation}
Now we need a subspace $S(\widetilde{\mathscr{H}}^{\mathrm{HS}})\subset S(\mathscr{H}^{\mathrm{HS}})$ which projects to $\mathcal{M}^{+}$ under $\pi$. We find this subspace to be 
\begin{equation}
S(\widetilde{\mathscr{H}}^{\mathrm{HS}})\equiv\bigl\{W\in
S(\mathscr{H}^{\mathrm{HS}})\,\big|\,\mathrm{Ker}(W)=0\bigr\}. 
\end{equation}
Among others, we just excluded the projective Hilbert space from $\mathcal{M}^+$. But it turns out that, once the metric tensor field we are interested in has been derived on $\mathcal{M}^+$, it can be extended to equip the entire quantum state space $\mathcal{M}$ with a Riemannian metric \cite{Uhlmann1995}. 

To obtain the decomposition into horizontal tangent spaces $H_{W}$ and vertical tangent spaces $V_{W}$, $T_{W}S(\widetilde{\mathscr{H}}^{\mathrm{HS}})=H_{W}\oplus V_{W}$, we can again introduce a connection to the $U(\mathscr{H})$ bundle $S(\widetilde{\mathscr{H}}^{\mathrm{HS}})\stackrel{\pi}\rightarrow\mathcal{M}^{+}$ using the scalar product on Hilbert-Schmidt space. Note that the tangent spaces $T_{W}S(\mathscr{H}^{\mathrm{HS}})$ at a point $W\in S(\mathscr{H}^{\mathrm{HS}})$ can be identified with subspaces of $\mathscr{H}^{\mathrm{HS}}$ due to the Hilbert-space property of $\mathscr{H}^{\mathrm{HS}}$. For the push forward of a vector $X\in T_{W}S(\widetilde{\mathscr{H}}^{\mathrm{HS}})$ we obtain 
\begin{equation}\label{pistarX.eq}
\pi_{*}(X)=WX^{\dagger}+XW^{\dagger}\,\in T_{WW^{\dagger}}\mathcal{M}^{+}.
\end{equation}
Since $\pi$ eliminates all the vertical directions, a vector $Y$ is vertical if
\begin{equation}\label{eq:10} 
WY^{\dagger}+YW^{\dagger}=0.
\end{equation}
For horizontality of $X$, we thus require $\langle X,Y\rangle_{\mathrm{HS}}=0$ to hold for all vertical vectors $Y \in V_{W}$. This leads to the condition 
\begin{equation}\label{eq:11} 
X^{\dagger}W - WX^{\dagger}=0.
\end{equation}
Note that $X$, being a tangent vector to the point $W$, can equivalently be written as $\frac{\mathrm{d}}{\mathrm{d}t}W(t)|_{t=0}$ for some curve $W(t), W(0)=W$. Then, following \cite{Dabrowski1990, Uhlmann1989}, one can show that the ansatz $\mathrm{d}W=GW$, with $G$ a hermitian matrix-valued 1-form, solves equation \eref{eq:11}. For the 1-form $\mathrm{d}\rho$, defined as the exterior derivative of the density matrix $\rho=WW^{\dagger}$, this translates into
\begin{equation}\label{eq:12}
\mathrm{d}\rho=\rho G + G \rho.
\end{equation}
If we now define a metric tensor field on the tangent spaces of $\mathcal{M}^{+}$ by taking, again, the real part of the scalar product and admitting only horizontal vectors as arguments, we obtain the so-called Bures metric 
\begin{equation}\label{eq:13} 
g(P_{1}, P_{2})\equiv\Re\bigl\{\langle X_{1}^{H},
X_{2}^{H}\rangle_{\mathrm{HS}}\bigr\}=\frac{1}{2}\mathrm{tr}\,\mathrm{d}\rho\otimes
G (P_{1}, P_{2}), 
\end{equation}
with $P_{1}=\pi_{*}(X_{1}^{H})$, $P_{2}=\pi_{*}(X_{2}^{H})$. Solving equation \eref{eq:12} for the matrix elements of $G$ and using a spectral resolution of the identity operator $\mathds{1}=\sum_n |\psi_n\rangle\langle\psi_n|$ in terms of the eigenvectors of the density operator $\rho$, one obtains 
\begin{equation}\label{eq:14} 
g = \frac{1}{2}\sum_{n,m}\frac{\langle\psi_n|\mathrm{d}\rho|\psi_m\rangle \otimes 
\langle\psi_m|\mathrm{d}\rho|\psi_n\rangle}{p_n + p_m}
\end{equation} 
for the metric tensor field, which was first found by H\"{u}bner \cite{Huebner1992}. The $p_n$ are the eigenvalues of the density operator $\rho$, which can be interpreted as statistical weights. In reference \cite{Zanardi2007}, by expanding $\mathrm{d}\rho$ in terms of the eigenstates of $\rho$, equation \eref{eq:14} is taken as a starting point for decomposing $g$ into two parts, 
\begin{equation}\label{eq:15}
g = g^{\mathrm{cl}} + g^{\mathrm{nc}},
\end{equation}
with 
\begin{equation}\label{eq:16}
g^{\mathrm{cl}} \equiv \frac{1}{4}\sum_n
\frac{1}{\sqrt{p_{n}}}\mathrm{d}p_{n}\otimes\frac{1}{\sqrt{p_{n}}}\mathrm{d}
p_{n} 
\end{equation}
and
\begin{equation}
g^{\mathrm{nc}} \equiv
\frac{1}{2}\sum_{n,m}\frac{(p_{n}-p_{m})^{2}}{p_{n} +
p_{m}}\left\langle
\psi_{n}|\mathrm{d}\psi_{m}\right\rangle\otimes\left\langle
\mathrm{d}\psi_{m}|\psi_{n}\right\rangle 
\label{eq:17}.
\end{equation}
The so-called classical ($\mathrm{cl}$) contribution $g^{\mathrm{cl}}$ formally coincides with the Fisher-Rao metric of classical information geometry \cite{Zanardi2007}. $g^{\mathrm{nc}}$, in contrast, was dubbed the non-classical ($\mathrm{nc}$) contribution. In reference \cite{Zanardi2007} it was also shown that the Bures metric indeed reduces to the Fubini-Study metric for pure states.

As a last step, we need to argue that the Bures metric defined on $\mathcal{M}^+$ can be extended to $\mathcal{M}$: An explicit calculation for finite systems reveals that the subspaces corresponding to zero-eigenvalues do not contribute to the trace operation which finally yields the distance between two density operators. Hence, equations \eref{eq:16} and \eref{eq:17} can be continued to $\mathcal{M}$ without modifications, and we have successfully constructed a fidelity metric on the quantum state space $\mathcal{M}$. The two expressions \eref{eq:16} and \eref{eq:17} form the starting point for our discussion of phase transitions at finite temperature and their relation to the Riemannian structure of quantum state space. 

\section{Isotropic LMG Model}\label{sec:4}

In this section we introduce the Lipkin-Meshkov-Glick (LMG) model in its spin formulation and give some of its basic properties, following mainly the presentation in \cite{Dusuel2005}. We then specialize to the isotropic case which is exactly solvable with little effort even in the case of finite systems and shortly report on its ground-state structure. Finally, the exact thermodynamic solution is recalled. 

The LMG model describes $N$ spin-$1/2$ degrees of freedom residing on the vertices of a graph. The spins interact through a ferromagnetic exchange coupling of infinite range, i.e., all spin pairs interact with equal strength, 
\begin{equation}\label{eq:18}
\mathcal{H}_{\mathrm{LMG}} = -\frac{1}{N}\sum_{i<j}\mvec{\sigma}_i^\dagger\mathcal{C}\mvec{\sigma}_j - h\sum_i \sigma_i^z, 
\end{equation}
where
\begin{equation}\label{eq:19}
\mvec{\sigma}_i=\left(\sigma_i^x,\sigma_i^y,\sigma_i^z\right)^{\mathrm{t}},\qquad
\mvec{\sigma}_i^\dagger=\left(\sigma_i^x,\sigma_i^y,\sigma_i^z\right).
\end{equation}
Here, $i,j$ label the graph vertices and $\mvec{\sigma}_i$ denotes the vector of Pauli matrices acting on the Hilbert subspace $\mathscr{H}_i\cong\mathds{C}^2$ corresponding to each vertex. The Pauli-vector components satisfy $[\sigma_i^\mu,\sigma_j^{\nu\vphantom{\mu}}] = 2\mathrm{i}\,\delta_{ij}\epsilon_{\mu\nu\kappa}\sigma_j^\kappa$, where we used Greek indices to label spatial vector components. The full Hilbert space is given by the tensor product
\begin{equation}
\mathscr{H}=\bigotimes_{i=1}^{N}\mathscr{H}_i\cong\left(\mathds{C}^2\right)^{\otimes N}.
\end{equation}
The coupling matrix $\mathcal{C}$ in \eref{eq:18} is given by $\mathcal{C}=\mathrm{diag}(1,\gamma,0)$, with anisotropy parameter $\gamma$. A factor of $1/N$ is included in \eref{eq:18} to ensure a finite free energy per degree of freedom when taking the thermodynamic limit. Moreover, an external magnetic field of strength $h$, pointing in the $z$-direction, tries to align the spins along this direction. 

The model dynamics can be formulated entirely in terms of the total spin
\begin{equation}
\mvec{S} = \frac{1}{2}\sum_{i=1}^{N}\mvec{\sigma}_i.
\end{equation}
Its components obey the usual angular momentum commutation relations $\left[S_{\mu},S_{\nu}\right] = \mathrm{i}\,\epsilon_{\mu\nu\kappa} S_{\kappa}$, yielding $\left[\mvec{S}^2, S_\mu\right] = 0$ for all $\mu$. Introducing spin-raising and -lowering operators $S_{\pm} = \left(S_{x} \pm \mathrm{i}S_{y}\right)/2$, the Hamiltonian can be rewritten as 
\begin{equation} 
\mathcal{H}_{\mathrm{LMG}} = -\frac{1+ \gamma}{N} \left(\mvec{S}^2 - S_z^2 - \frac{N}{2}\right) 
- 2 h S_z - \frac{1-\gamma}{2 N}\left(S_{+}^{2} + S_{-}^{2}\right) \label{eq:20}.
\end{equation} 
Since $\left[\mvec{S}^2,\mathcal{H}_{\mathrm{LMG}}\right]=0$, $\mvec{S}^2$ is a conserved quantity under the dynamics induced by $\mathcal{H}_{\mathrm{LMG}}$. Thus, the Hilbert space can be decomposed as
\begin{equation}
\mathscr{H}\cong \left(\mathds{C}^{2}\right)^{\otimes N} \cong \bigoplus_{S}d_{S}\,\mathcal{D}_{S},
\end{equation}
where $d_{S}$ denote the multiplicities of irreducible and unitary $SU(2)$-representations $\mathcal{D}_{S}$ of dimension $\dim\mathcal{D}_{S}=2S+1$. For convenience, we choose $N$ even in the following, obtaining $S\in\{0,\dots,N/2\}$. Moreover, the Hamiltonian is invariant under time reversal $(h\mapsto -h, \mvec{\sigma} \mapsto -\mvec{\sigma})$. Therefore, all eigenvalues are at least twice degenerate (Kramers degeneracy), $E_{n}(h)=E_{n^{\prime}}(-h)$, where $n$ and $n^{\prime}$ denote distinct sets of quantum numbers. Due to this symmetry, we can restrict the discussion of the spectral properties of $\mathcal{H}_{\mathrm{LMG}}$ to the case $h\geqslant 0$. We now specialize to the isotropic model, and comment on the anisotropic case in section \ref{sec:7}. The Hamiltonian \eref{eq:20} reduces in the isotropic case $\gamma=1$ to 
\begin{equation}\label{eq:21}
\mathcal{H}_{\mathrm{LMG}}^{\mathrm{iso}} = -\frac{2}{N}\left(\mvec{S}^2 - S_z^2 - \frac{N}{2}\right) - 2 h S_z.
\end{equation}
Since $\left[S_{z},\mathcal{H}_{\mathrm{LMG}}^{\mathrm{iso}}\right] = 0$, now also $S_{z}$ is an integral of motion. We denote by $|SM\rangle$ the simultaneous eigenstates of $S^2$ and $S_z$, where
\begin{equation}
\mvec{S}^2|SM\rangle=S(S+1)|SM\rangle,\qquad S_z|SM\rangle=M|SM\rangle.
\end{equation}
For every spin sector $\mathcal{D}_{S}$, the angular momentum eigenstates $|SM\rangle$, $M=-S,\dots,+S$, are eigenstates of the Hamiltonian, and $\mathcal{H}_{\mathrm{LMG}}^{\mathrm{iso}}$ is invariant under rotations about the $z$-axis. Its eigenvalues are given by 
\begin{equation}\label{eq:22}
E_{SM} = -\frac{2}{N}\left(S\left(S+1\right)-M^{2} - \frac{N}{2}\right) - 2 h M.
\end{equation}
$E_{SM}$ attains its minimum in the maximum-spin sector (i.e., for quantum number $S_{0}=N/2$) with magnetic quantum number
\begin{equation}\label{eq:23}
M_{0}=\cases{
\mathcal{I}(hN/2) & \mbox{for $0\leqslant h < 1$},\\
N/2 & \mbox{for $h\geqslant 1$},
}
\end{equation}
where
\begin{equation}\label{eq:24}
\mathcal{I}\left(x\right)=\cases{
 \lfloor x \rfloor & \mbox{for $x = \lfloor x \rfloor + \delta$, $\delta \in [0,1/2)$},\\
 \lceil x \rceil & \mbox{for $x = \lfloor x \rfloor + \delta$, $\delta \in [1/2,1)$},
}
\end{equation} 
is the rounding function. The value of the external field $h$ therefore determines which of the angular momentum states $\in\mathcal{D}_{N/2}$ is selected as the ground state. At certain values of $h$ the ground state switches from one $M$-value to another (see equation \ref{eq:23}), and these points of degeneracy are termed level crossings. In the thermodynamic limit $N\to\infty$, the ground-state energy per spin converges towards a continuous function of $h$ \cite{Dusuel2005}, being infinitely differentiable almost everywhere. Only at $h=\hc|_{T=0}\equiv\pm 1$ its second derivative with respect to $h$ is discontinuous, signaling the above-mentioned phase transitions. 

We now recall the exact thermodynamic solution of the LMG model, which is a special case of a result by Pearce and Thompson \cite{Pearce1975} obtained for a large class of mean-field type spin models in an external field. For the isotropic LMG model, the free energy per spin in the thermodynamic limit $N\rightarrow\infty$ is given by 
\begin{equation}\label{eq:25}
\fl f(\beta,h) \equiv
-\lim_{N\to\infty}\frac{1}{N\beta}\ln\mbox{tr}\exp(-\beta
\mathcal{H}_{\mathrm{LMG}}^{\mathrm{iso}}) 
= \frac{1}{2}\mu_{xy}^{2} - \beta^{-1}\ln{\left(2\cosh{\left(\beta\sqrt{\mu_{xy}^{2} + h^2}\right)}\right)},
\end{equation} 
where $\beta$ denotes inverse temperature. The relative magnetization in $z$-direction, $\mu_{z}=-\partial f/\partial h$, is completely determined by the value of the external field $h$ and the scalar order parameter $\mu_{xy}=\mu_{xy}(\beta,h)$. The latter obeys the self-consistency equation
\begin{equation}\label{eq:27}
\mu_{xy}^2 + h^2 = \biggl(\tanh\left(\beta\sqrt{\mu_{xy}^2 + h^2}\right)\biggr)^{2} 
\end{equation} 
and has the interpretation of a relative in-plane magnetization with respect to the maximum total spin $N/2$,
\begin{equation}\label{eq:26}
\mu_{xy}^2= \lim_{N\rightarrow\infty}\frac{2}{N^{2}}\left\langle\biggl(\sum_{i}\sigma_i^x\biggr)^2 + \biggl(\sum_i\sigma_i^y\biggr)^2\right\rangle 
\end{equation}
where $\langle\cdot\rangle$ denotes a thermal-equilibrium average.

\begin{figure}[t]\center
\includegraphics[width=6.8cm]{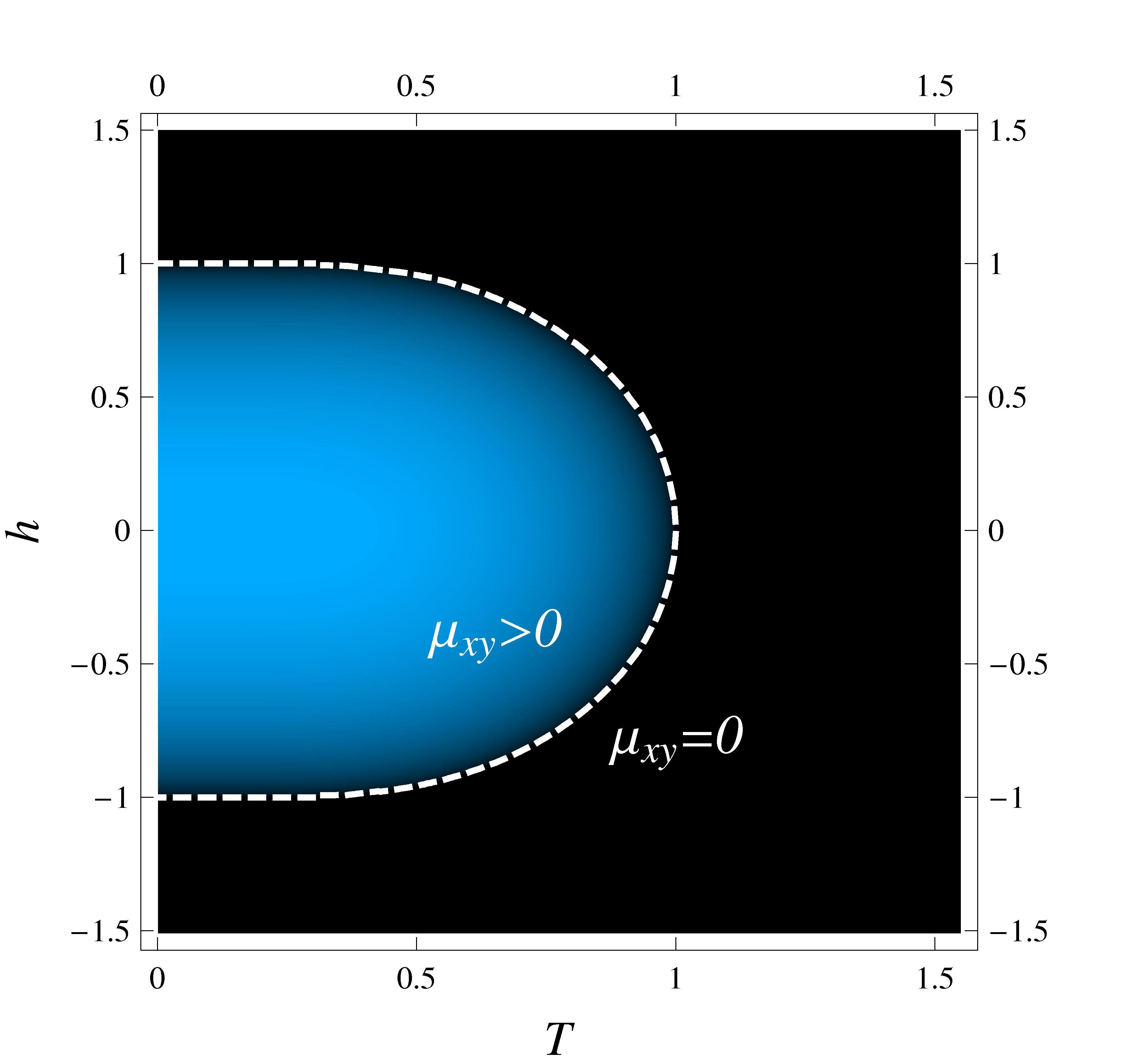}
\caption{\label{fig:phase} Phase diagram of the isotropic LMG model plotted in the $(T,h)$-plane. The blue shaded area marks the ordered phase with finite in-plane magnetization $\mu_{xy}\neq0$, separated from the paramagnetic phase $\mu_{xy}=0$ (shown in black) by a line of phase transitions making up the phase boundary (dashed white).} 
\end{figure}

The self-consistency equation (\ref{eq:27}) determines the phase diagram completely, see figure \ref{fig:phase}. For fields with $|h|<1$ and temperatures $T$ below the critical temperature $T_{\mathrm{c}}(h)$, $\mu_{xy}$ takes non-zero values and vanishes continuously when approaching the phase boundary by either an increase in temperature or magnetic field. The phase boundary reached as $\mu_{xy}=0$ consists of all points $(\betac,\hc)$ that obey $\hc=\tanh(\betac \hc)$ or 
\begin{equation}\label{eq:28}
\betac(\hc)=\hc^{-1}\mathrm{arctanh}(\hc),
\end{equation}
where $\betac=1/T_{\mathrm{c}}$ and $\hc$ denote critical values of inverse temperature and magnetic field. In summary, the LMG model shows a ferromagnetically-ordered phase separated from a paramagnetic phase by a line of continuous phase transitions. All these exact thermodynamic results for the infinite system coincide with results obtained from a mean-field treatment as reported in \cite{Quan2008} and \cite{Sollich2000}. 

\section{Metric Tensor Field for Thermal States}\label{sec:5}

In this section we compute the metric tensor field on the submanifold of thermal states. A thermal equilibrium state (or Gibbs state) of the isotropic LMG model is given by 
\begin{equation}\label{thermalstate}
\rho=\frac{1}{\mathcal{Z}_N(\beta,h)}\exp\left(-\beta\mathcal{H}_{\mathrm{LMG}}^{\mathrm{iso}}\right),
\end{equation}
where
\begin{equation}
\mathcal{Z}_{N}(\beta,h) =\sum_{S=0}^{N/2}\,d_{S}\sum_{M=-S}^{+S} \left\langle SM\right|
\mathrm{e}^{-\beta\bigl(-\frac{2}{N}\left(\mvec{S}^2 - S_z^2 - \frac{N}{2}\right) - 2 h S_z
\bigr)}\left|SM\right\rangle
\end{equation}
is the canonical partition function. Here, the density operator inherits a dependence on $\beta$ and $h$ from the Hamiltonian and the partition function. 

For finite systems, equation \eref{thermalstate} defines a parameterization of the submanifold $\mathcal{G}$ of thermal states. Equivalently, we can take this as a trivial chart $\rho(\beta,h)\mapsto(\beta,h)$, defining local coordinates on $\mathcal{G}$. Vector fields (and, analogously, 1-forms or higher rank tensor fields) can then be expressed with respect to the coordinate basis $\{\partial_{\beta}, \partial_{h}\}$. 

As a first step, we use the decomposition $\mathscr{H}\cong \bigoplus_{S}d_{S}\,\mathcal{D}_{S}$ of the Hilbert space of the LMG model to cast the spectral representation of $g$ [equations \eref{eq:15}--\eref{eq:17}] in a different form: We solve equation \eref{eq:12}, separately within every spin sector $\mathcal{D}_{S}$, for the matrix elements of $G$ with respect to the $(2S+1)$-dimensional basis $\{|SM\rangle\}$. Plugging the result into equation \eref{eq:13} and taking the trace, we arrive at the expressions 
\begin{equation}\label{eq:30}
g^{\mathrm{cl}} = \frac{1}{4}\sum_{S} d_{S}\sum_{M} \frac{1}{\sqrt{p_{SM}}}\mathrm{d}p_{SM}\otimes\frac{1}{\sqrt{p_{SM}}}\mathrm{d} p_{SM}
\end{equation}
and
\begin{equation}
\fl g^{\mathrm{nc}} =
\frac{1}{2}\sum_{S}d_{S}\sum_{M,M^{\prime}}\frac{\left(p_{SM}-p_{SM^{\prime}}\right)^2}{p_{SM}
+ p_{SM^{\prime}}}\left\langle
SM|\mathrm{d}|SM^{\prime}\right\rangle\otimes\left\langle
SM|\mathrm{d}|SM^{\prime}\right\rangle^{*} 
=0 \label{eq:31}.
\end{equation}
Here, by
\begin{equation}
p_{SM}=\frac{1}{\mathcal{Z}_N}\exp(-\beta E_{SM})
\end{equation}
we denote the statistical weights with energies $E_{SM}$ as given in \eref{eq:22}. Since the eigenstates of $\mathcal{H}_{\mathrm{LMG}}^{\mathrm{iso}}$ do not carry any explicit $h$-dependence (nor, of course, any $\beta$-dependence), the non-classical contribution $g^{\mathrm{nc}}$ vanishes by virtue of $\rmd\ket{SM}=0$, and we are left with the classical Fisher-Rao contribution \eref{eq:30}. This is maybe not too surprising: Since the operators $\mvec{S}^2$, $S_z$ in the Hamiltonian of the isotropic LMG model are commuting, a (classical) probability distribution can be assigned, and the corresponding information geometrical metric is known to be the one of Fisher-Rao. A more detailed and more general discussion of Hamiltonians consisting of commuting summands and the implications on the fidelity can be found in reference \cite{PaunkovicVieira08}.

It is straightforward to compute the 1-forms $\mathrm{d}p_{SM}$ in equation \eref{eq:30} with respect to the dual coordinate basis $\mathrm{d}\beta,$ $\mathrm{d}h$, yielding 
\begin{equation}
\mathrm{d}p_{SM}= \partial_{\beta}p_{SM}\mathrm{d}\beta + \partial_{h}p_{SM}\mathrm{d}h,
\end{equation}
where the partial derivatives can be rewritten as
\numparts
\begin{eqnarray}
\partial_{\beta}p_{SM} &= p_{SM}\left(\left\langle\mathcal{H}_{\mathrm{LMG}}^{\mathrm{iso}}\right\rangle - E_{SM}\right),\label{eq:32}\\
\partial_{h}p_{SM} &= -2\beta p_{SM}\left(\left\langle S_{z}\right\rangle - M\right).\label{eq:33}
\end{eqnarray}
\endnumparts
Expanding the metric tensor field with respect to the rank two tensor basis $\mathrm{d}\beta\otimes\mathrm{d}\beta$, $\mathrm{d}\beta\otimes\mathrm{d}h$, etc., we obtain 
\numparts
\begin{eqnarray}
g_{\beta\beta} 
& \equiv &\frac{1}{4}
\sum_S d_S\sum_M \frac{\left(\partial_\beta p_{SM}\right)^2}{p_{SM}}
 ,\\
g_{hh} 
&\equiv&\frac{1}{4}
\sum_S d_S\sum_M \frac{\left(\partial_h p_{SM}\right)^2}{p_{SM}}
,\\
g_{h\beta} 
&\equiv&\frac{1}{4}
\sum_S d_S\sum_M \frac{\partial_\beta p_{SM}\partial_h p_{SM}}{p_{SM}}
,
\end{eqnarray}
\endnumparts
and, furthermore, $g_{\beta h}=g_{h\beta}$. Inserting (\ref{eq:32}) and (\ref{eq:33}) into the above equations, we find the metric components to be given by equilibrium fluctuations and correlations, 
\numparts
\begin{eqnarray}
g_{\beta\beta} & =
&\frac{1}{4}\left(\left\langle{\mathcal{H}_{\mathrm{LMG}}^{\mathrm{iso}}}^{2}\right\rangle
-
\bigl\langle\mathcal{H}_{\mathrm{LMG}}^{\mathrm{iso}}\bigr\rangle^{2}\right),
\label{eq:34a}\\ 
g_{hh} & = &\beta^{2}\left(\bigl\langle S_{z}^{2}\bigr\rangle - \bigl\langle S_{z}\bigr\rangle^{2}\right),\label{eq:34b}\\
g_{h\beta} & = & -
\frac{\beta}{2}\Bigl(\bigl\langle\mathcal{H}_{\mathrm{LMG}}^{\mathrm{iso}}
S_{z}\bigr\rangle -
\bigl\langle\mathcal{H}_{\mathrm{LMG}}^{\mathrm{iso}}\bigr\rangle\bigl\langle
S_{z} \bigr\rangle\Bigr).\label{eq:34c} 
\end{eqnarray}
\endnumparts
Analogous results have been derived in \cite{Zanardi2007} for the case of a quantum Ising spin chain. Since $\mathcal{H}_{\mathrm{LMG}}^{\mathrm{iso}}$ contains only commuting operators, the connected correlation functions on the right hand sides of \eref{eq:34a}--\eref{eq:34c} can, up to some pre-factors, be expressed as derivatives of the free energy \eref{eq:25}, yielding 
\numparts
\begin{eqnarray}
g_{\beta\beta} & = & -\frac{N}{4}\partial_{\beta}^{2}\left(\beta f\right), \label{eq:37a}\\
 g_{hh} & = & -\frac{N\beta}{4}\partial_{h}^{2}f, \\
g_{h\beta} & = & -\frac{N\beta}{4}\partial_{h}\partial_{\beta}f.\label{eq:37c}
\end{eqnarray}
\endnumparts
Note that these expressions suggest a close relationship to Ruppeiner geometry \cite{Ruppeiner1995}, where a Riemannian metric is \textit{defined} in terms of derivatives of a suitable thermodynamic potential, e.g.\ the free energy. In the case of a vanishing non-classical part $g^{\mathrm{nc}}$, the Bures metric for thermal states and the Ruppeiner metric indeed differ just by a coordinate transformation. 

The exact result for the free energy per spin \eref{eq:25} and the self-consistency condition \eref{eq:26} can now be used to compute the metric per spin in the thermodynamic limit, $g^{(\infty)}\equiv\lim_{N\rightarrow\infty}\frac{1}{N}g$. One would expect from equations \eref{eq:37a}--\eref{eq:37c} that, in this limit, the metric inherits a nonanalytic behaviour from the nonanalyticity of the free energy $f$, but we will see in the following that the situation is even a bit more involved.

\subsection{Ordered phase}

\begin{figure}\center
\includegraphics[width=6.8cm]{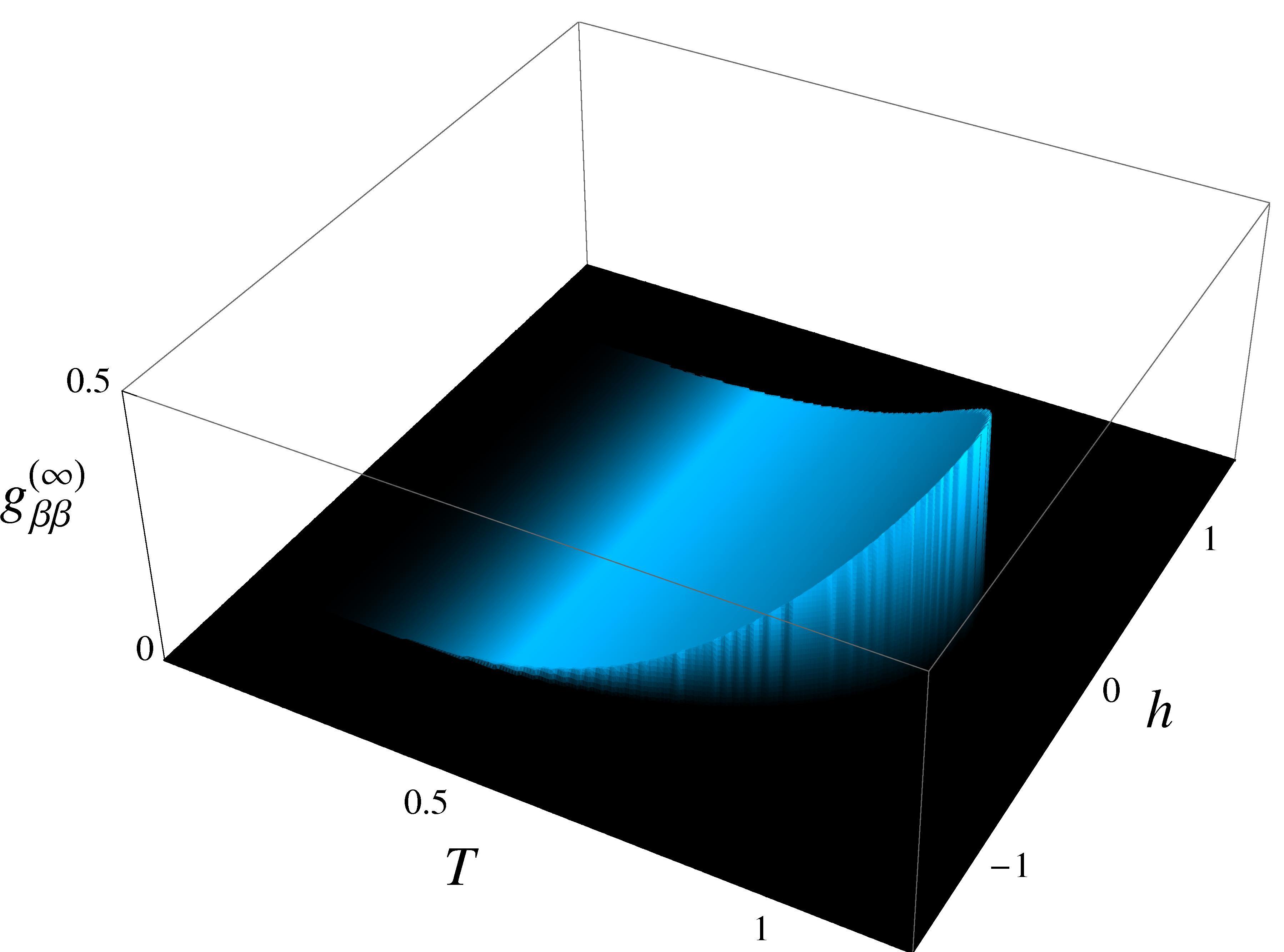}
\hspace{8mm}
\includegraphics[width=6.8cm]{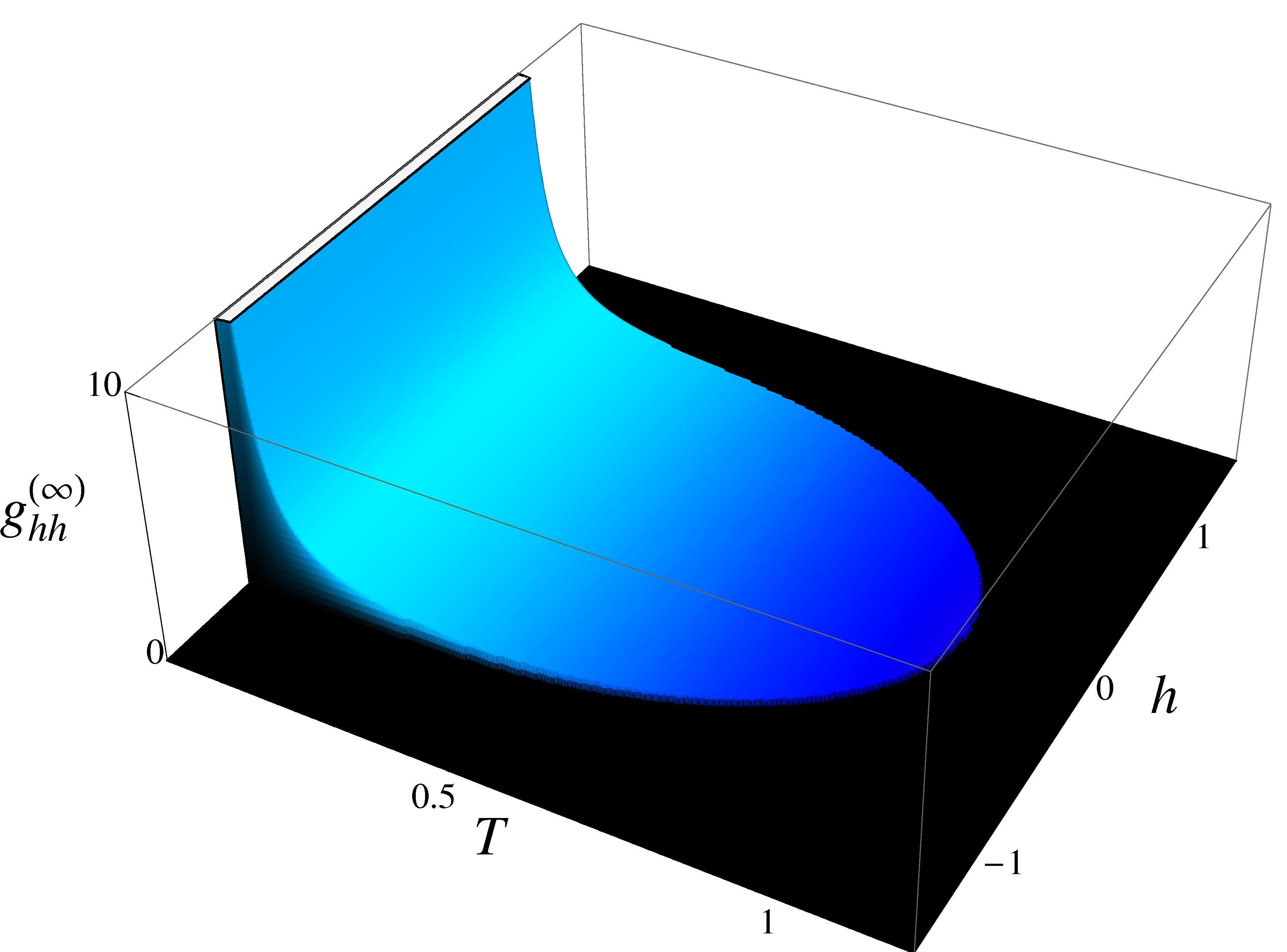}
\caption{\label{fig:1} Components $g_{\beta\beta}^{(\infty)}$ and $g_{hh}^{(\infty)}$ of the metric in the thermodynamic limit as functions of temperature $T$ and external magnetic field $h$. When approaching the phase boundary, $g_{\beta\beta}^{(\infty)}$ increases, suggesting enhanced distinguishability under variations of temperature. From the divergence of $g_{hh}^{(\infty)}$ at $T=0$ one can see that the Bures metric is not well-defined for the ground state in the thermodynamic limit (nor is it well-defined for finite systems at $T=0$).} 
\end{figure}

For the ordered phase with $\mu_{xy}\neq 0$ we can collect the metric components in the form of a diagonal matrix,
\begin{equation}\label{eq:38}
\left( 
\begin{array}{cc}
g_{\beta\beta}^{(\infty)} & g_{\beta h}^{(\infty)} \\
g_{h \beta}^{(\infty)} & g_{hh}^{(\infty)}
\end{array}
\right)
=
\frac{1}{4}\left(
\begin{array}{cc}
\frac{\left(\mu_{xy}^{2} + h^2\right)\left(1 + \mu_{xy}^2 +
 h^2\right)}{1 - \beta\left[1 - \left(\mu_{xy}^{2} +
 h^2\right)\right]} 
& 0\\
0 & \beta
\end{array}
\right).
\end{equation}
Equation \eref{eq:38} provides the components of a well-defined Riemannian metric, and we can now interpret the metric components as indicators of how well thermal states with close-by values of $\beta$ and $h$ can be distinguished. The graphs of $g_{\beta\beta}^{(\infty)}$ and $g_{hh}^{(\infty)}$ in the $(T,h)$-plane are shown in figure \ref{fig:1}. 

\begin{figure}\center
\includegraphics[width=7cm]{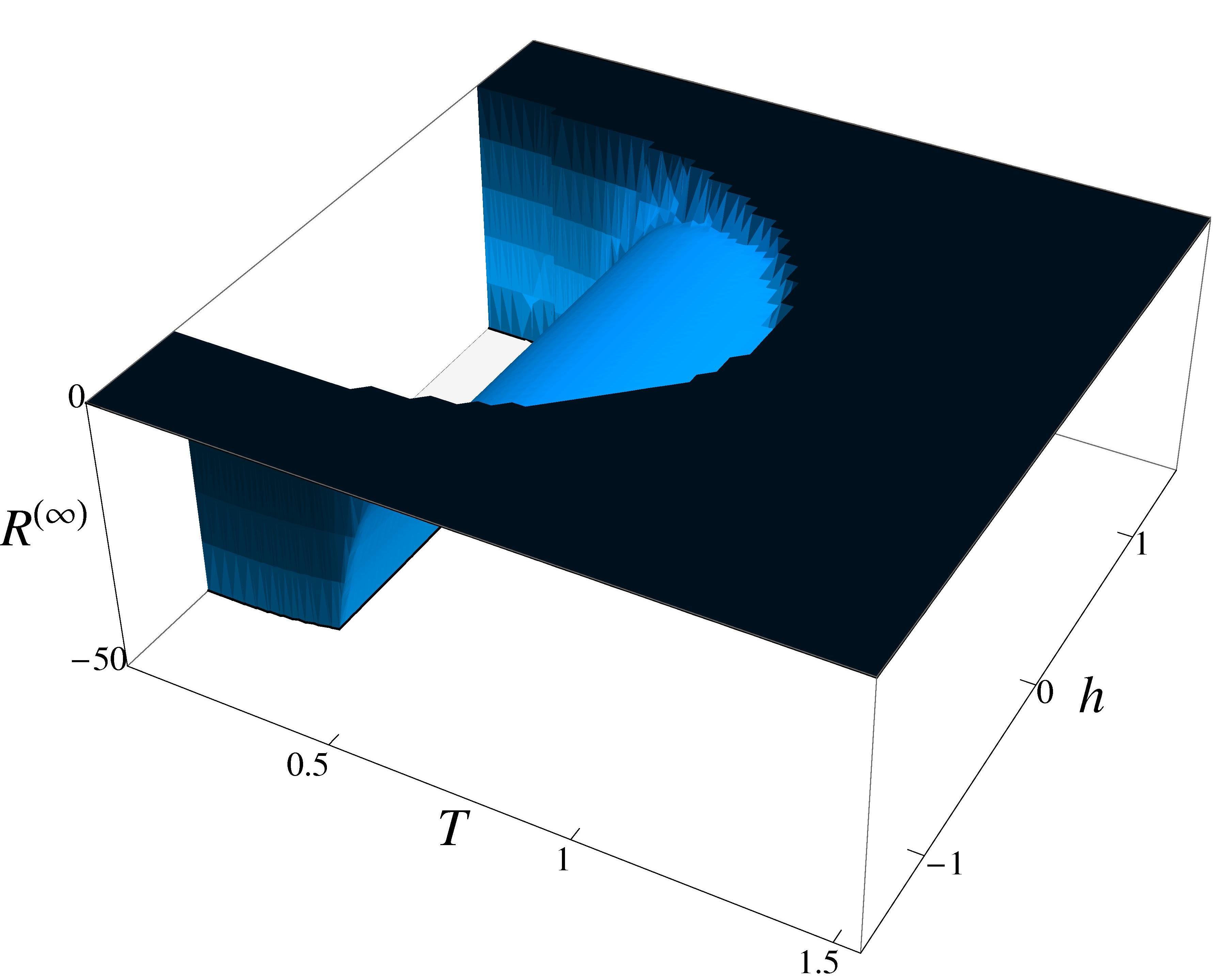}
\hspace{8mm}
\includegraphics[width=6cm]{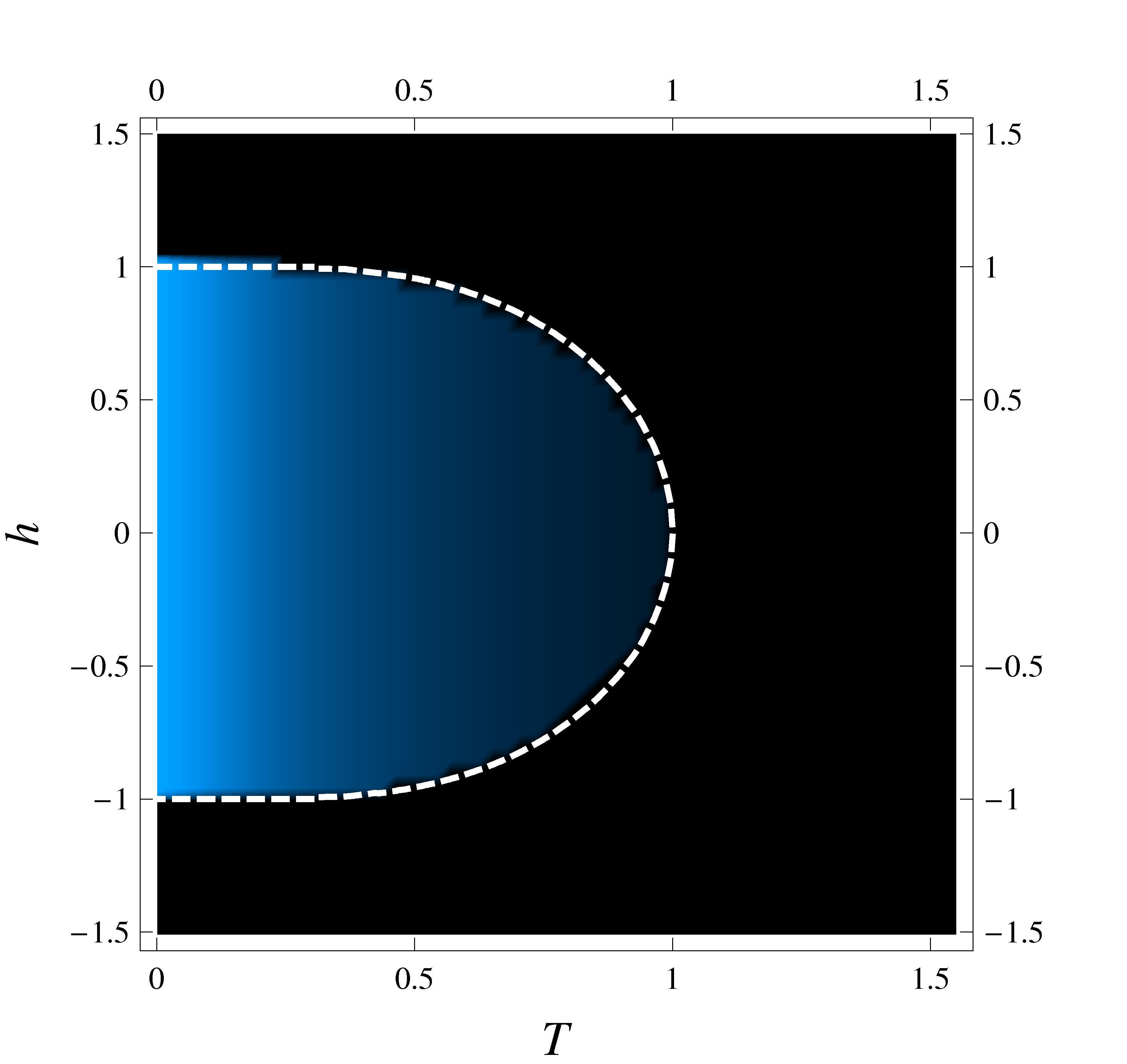}
\caption{\label{fig:ricci} 
Ricci scalar $\mathcal{R}^{(\infty)}$ of the isotropic LMG model in the thermodynamic limit as functions of temperature $T$ and external magnetic field $h$. As for the metric components in figure \ref{fig:1}, $\mathcal{R}^{(\infty)}$ is well-defined only for the ferromagnetically ordered phase, and the breakdown of its existence can be interpreted as a signal of the phase transition. Note that, where defined, $\mathcal{R}^{(\infty)}$ is negative, suggesting a classical-type behaviour of the system.} 
\end{figure}

Instead of considering the metric components separately, one can combine them to compute the Ricci scalar, a quantity characterizing the curvature of a manifold. We have computed the Ricci scalar by making use of the Maurer-Cartan \cite{MNakahara} equations. Here, we skip the details of this calculation and present only the final result in the thermodynamic limit, 
\begin{equation}
\mathcal{R}^{(\infty)} = \frac{2}{PQ} \left( \frac{(\partial_\beta P)( \partial_h Q)}{Q^2}-\frac{\partial_\beta^2 P}{Q}-\frac{\partial_h^2 Q}{P}\right),
\end{equation}
where $P=\sqrt{\beta}/2$ and $Q=\sqrt{\mu_{xy}\partial_\beta \mu_{xy}}/2$. The graph of $\mathcal{R}^{(\infty)}$ is shown in figure \ref{fig:ricci}. We observe that $\mathcal{R}^{(\infty)}$ is negative in the entire ordered phase. In reference \cite{Zanardi2007}, it was conjectured that a negative Ricci scalar should correspond to the ``classical realm'' of a given system. Since the Hamiltonian of the isotropic LMG model contains only mutually commuting terms, it may be regarded as classical, and the negative Ricci scalar we observe is in agreement with the conjecture.

\subsection{Paramagnetic phase} 

Surprisingly at first sight, the metric not only becomes singular at the phase boundary, but changes its structure entirely from one phase to the other. Writing the rescaled metric tensor in matrix form,
\begin{equation}
\left(\begin{array}{cc} g^{(\infty)}_{\beta\beta} & g^{(\infty)}_{\beta h} \\ g^{(\infty)}_{h \beta} & g^{(\infty)}_{hh}\end{array}\right)
=
\frac{1}{4}\Bigl(\cosh{(\beta h)}\Bigr)^{-2}
\left(
\begin{array}{cc}
 h^{2} & h \beta \\
 h \beta & \beta^{2}
\end{array}
\right),
\end{equation}
we find that in the disordered phase with $\mu_{xy}=0$, the matrix of the metric components has vanishing determinant. This in turn implies that in the disordered phase the rank two tensor becomes degenerate and is not a proper Riemannian metric anymore. Since we started out with a Riemannian metric in the finite-system case, the limit $N\to\infty$ must have destroyed this property.

One can understand the physical origin of this effect by considering expression \eref{eq:25} for the free energy of the LMG model: For vanishing in-plane magnetization $\mu_{xy}=0$, the free energy is identical to that of a spin system coupled to an external field $h$, but without any spin-spin interaction whatsoever. Such a system is governed by the Zeeman-Hamiltonian $\mathcal{H}=-h S_{z}$, and the corresponding thermal state is given by $\rho=\mathcal{Z}^{-1}\exp(\beta hS_{z})$ with partition function $\mathcal{Z}=\mathrm{tr}\exp(\beta hS_{z})$. For this system, all $(\beta,h)$ with $\beta h=\mathrm{const}.$ parametrize the same density operator. The very same situation occurs also for the paramagnetic phase with $\mu_{xy}=0$ of the LMG model in the thermodynamic limit, leading to the mentioned degeneracy of the matrix of the metric components. As a consequence, thermal states of this phase should be parametrized by only a single parameter, namely the reduced field $\bar{h}\equiv\beta h$. The corresponding metric on such a one-dimensional manifold can be obtained by a computation similar to the two-dimensional case reported above, yielding 
\begin{equation}\label{eq:39}
g^{(\infty)} = \left(4\cosh\bar{h}\right)^{-1}\mathrm{d}\bar{h}\otimes\mathrm{d}\bar{h}.
\end{equation} 

\section{Fubini-Study Limit}\label{sec:6}

We have mentioned at the end of section \ref{sec:3} that the Bures metric reduces to the Fubini-Study metric when considering pure states. This should in principle allow us to define a Fubini-Study metric on the ground-state manifold parametrized by a ground-state mapping $h\mapsto|\psi_{GS}\rangle$. However, we have observed in section \ref{sec:4} that the ground state of the isotropic LMG model is the angular momentum eigenstate $\ket{N/2,M_0(h)}$ where, according to equation \eref{eq:23}, $M_0(h)$ is selected by the rounding function $\mathcal{I}$. As a consequence, no differentiable parametrization of the ground state exists. Since this property naturally carries over to the respective chart mapping, the Fubini-Study metric on the ground-state manifold of the isotropic LMG model in the finite system is not well defined. Similar issues, related to ground state level crossings and their effect on the fidelity, are also discussed in reference \cite{PaunkovicVieira08}.

In order to investigate the ground state behaviour in the infinite system, we can alternatively study the limit $T\to0$ of the Riemannian metric characterized by \eref{eq:38}. The component $g_{\beta\beta}^{(\infty)}$ is found to vanish in this limit, in agreement with the fact that, according to equation \eref{eq:34a}, it is proportional to the specific heat. The component $g_{hh}^{(\infty)}$, however, diverges for $T\to 0$ asymptotically as $T^{-1}$. Hence the (rescaled) ground-state metric is not well defined in this limit, although the ground state energy becomes a continuous function of the external field $h$ in the thermodynamic limit. 

\section{Remarks on the Anisotropic Case}\label{sec:7}

The problems we encountered in the previous section when trying to compute a metric on the ground-state manifold of the isotropic LMG model can be traced back to the fact that the Hamiltonian consists only of mutually commuting terms, which in turn allows for level crossings. For the anisotropic case, in contrast, we would expect a well-defined ground state metric. Exact results for the spectrum and eigenvalues, exist also for the anisotropic LMG model \cite{Pan1999, Morita2006}. In principle, these results would allow one to compute the metric on the manifold of thermal states, but unfortunately, they are expressed as rather complicated multiple sums, with coefficients given as solutions of differential equations, which makes the calculation quite difficult in practice. A more accessible result is given in reference \cite{Ribeiro_etal2007,Ribeiro_etal2008} for the energy spectrum of this model in the thermodynamic limit but, as far as we can see, this might not be sufficient for a computation of the fidelity metric.
 
Alternatively, one might try to compute the metric in mean-field approximation. Knowing that, in the thermodynamic limit, the mean-field solution of the (isotropic or anisotropic) LMG model coincides with the exact solution, one might hope to obtain the exact metric from a mean-field calculation as well. Unfortunately this is not the case, nor are we aware of any other approximation that retains enough of the original quantum state space structure in order to deliver an accurate description of the underlying geometry.

In contrast to the isotropic LMG model where only the classical part \eref{eq:30} was found to contribute to the metric, we expect the anisotropic case to have a non-zero non-classical contribution \eref{eq:31}. Furthermore, it will not anymore be possible to completely express the metric in terms of derivatives of the free energy, and only then the characterization of phase transitions by means of the fidelity metric would really go beyond a thermodynamic description. 

\section{Conclusions}\label{sec:8}

In this paper we have followed the idea that a suitable metric on quantum state space can be used to identify and characterize both 
classical and quantum phase transitions. We have reviewed how such a fidelity metric is constructed, either on the space of pure states $\mathcal{P}(\mathscr{H})$ or on the space of state operators $\mathcal{M}$. From the Bures metric, i.e., the fidelity metric on $\mathcal{M}$, the metric tensor field on the submanifold of thermal states has been derived. As an application of these concepts, we studied the LMG model of spin-$1/2$ degrees of freedom sitting on the vertices of a fully connected graph. The choice of this model was mainly motivated by the fact that its thermodynamics is particularly simple to solve, and exact results are available for the free energy per spin in the thermodynamic limit, both for the isotropic and the anisotropic LMG model. 

For the isotropic LMG model, we computed the metric tensor field on the submanifold of thermal states, and we found that all metric components can be written as derivatives of the free energy. This implies a close relation to Ruppeiner geometry, but this should be a peculiarity of models with purely classical contributions to the metric. Another peculiar feature special to the isotropic case is that on the ground-state manifold the metric is not well defined, neither by direct construction from the finite system nor by a detour via thermal states and the subsequent zero-temperature limit. This can be seen as a consequence of level crossings which occur in this case, but are avoided in the anisotropic model. 

As expected, we find that the phase transition of the isotropic LMG model occurring at the transition line \eref{eq:28} in the $(T,h)$-plane is well captured by the metric components. In a way, this signature is even more pronounced than for other models which had been studied before: Not only do the metric components show a singularity or discontinuity, but we find that, in the thermodynamic limit, the tensor field on the $(T,h)$-plane becomes degenerate for the paramagnetic phase and therefore ceases to be a proper Riemannian metric. 

It would be worthwhile to compare these results to the corresponding metric of the anisotropic LMG model. Here, we expect the metric to be well-defined on the ground-state manifold, and the non-classical part \eref{eq:31} of the metric to give a non-vanishing contribution.

\section*{Acknowledgments}
Most of the work reported was done while D.~D.~S.\ and M.~K.\ were affiliated with the Uni\-ver\-si\-t\"at Bay\-reuth.\\

\bibliographystyle{unsrt}
\bibliography{FidelityLMG}

\end{document}